# From surface Fermi arcs to Fermi loops in the Dirac semimetal $Cd_3As_2$

An-Qi Wang[1]*, Tong-Yang Zhao[1], Chuan Li[2]*, Alexander Brinkman[2], Chun-Guang Chu[1] and Zhi-Min Liao[1,3]*

[1] State Key Laboratory for Mesoscopic Physics and Frontiers Science Center for Nano-optoelectronics, School of Physics, Peking University; Beijing 100871, China.

[2] MESA+ Institute for Nanotechnology, University of Twente, 7500 AE Enschede, The Netherlands

[3] Hefei National Laboratory, Hefei 230088, China.

*Email: anqi0112@pku.edu.cn; chuan.li@utwente.nl; liaozm@pku.edu.cn

**Abstract:**

Arc-like topological surface states, *i.e.*, surface Fermi arcs, have long been recognized as the hallmark of Dirac semimetals. However, recent theories suggest that the surface Fermi arcs could evolve into closed Fermi loops, akin to surface states in topological insulators, while preserving the bulk Dirac semimetal phase. Here we experimentally reveal the evolution of Fermi arcs to Fermi loops in the surface-modified Dirac semimetal $Cd_3As_2$ nanoplate through gate voltage-dependent spin transport and quantum oscillation measurements. Surface modification, achieved by heavy metal atom deposition and water molecule adsorption, leads to an increase in the current-induced spin polarization at higher gate voltages, contrasting with the decrease observed in the pristine nanoplate. We also observe surface Shubnikov-de Haas oscillations with frequencies that scale linearly with gate voltage, aligning with a Fermi loop scenario. These findings indicate a transition from Fermi arcs to a closed Fermi loop in the surface-modified $Cd_3As_2$ nanoplate, consistent with the theoretically predicted fragile topological nature of $Cd_3As_2$. Our research offers profound insights into the transitions among these subtle topological states in Dirac semimetals, paving the way for manipulating topological surface states for high-performance spintronic devices.

Dirac semimetals (DSMs) have sparked enormous interest due to their bulk Dirac fermions and surface Fermi arcs, both of which exhibit exotic physical properties attractive for topological electronics [1-5]. DSM features with linear energy-momentum dispersion in three dimensions, and its conduction and valence band intersect at discrete points, called Dirac points. Dirac semimetals such as $Cd_3As_2$ and $Na_3Bi$ have been theoretically proposed [6,7] and experimentally identified by optical spectroscopy [8-11], scanning tunneling spectroscopy [12] and magneto-transport measurements [13-16]. In the presence of time-reversal symmetry and inversion symmetry, each Dirac node is described by a four-component relativistic Dirac equation, which can be viewed as overlapping of two Weyl nodes with opposite chiralities. Additional crystalline symmetry such as rotational symmetry is needed to prohibit the hybridization of the two Weyl nodes and stabilize the presence of Dirac points [7,17,18]. When these symmetries are broken by perturbations, the Dirac semimetal can transform into other topological phases [19-29]. For example, upon breaking the time-reversal symmetry or inversion symmetry, a Dirac point would split into two Weyl nodes, leading to the emergence of Weyl semimetal phase [19]. Applying external strains or pressures that affect the crystalline symmetry can open gap at the Dirac point, resulting in a phase transition to a topological insulator or even a superconductor [26-29].

Another notable feature of DSMs is the presence of surface Fermi arcs [4,5,30-33], which connect the surface projections of two Weyl points [34-40]. In contrast to the open surface Fermi arcs in Weyl semimetals (WSMs), DSMs exhibit a pair of surface Fermi arcs [3,6,7,41], due to each Dirac node comprising two Weyl nodes with opposite chiralities [Figs. 1(a) and 1(b)]. The arc-like surface states have led to many intriguing transport phenomena in DSMs, such as $\pi$ Aharonov−Bohm (AB) effect [42,43], current-induced spin polarization [44-46], enhanced Josephson supercurrent [47,48], magnetoresistance oscillations [49-53] and quantum Hall effect [20,54-58]. Recent theoretical calculations suggest that the arc-like surface states could transit into Fermi loop surface states without triggering any bulk phase transitions [59-61]. When a symmetry-breaking surface potential is applied, these arc-like surface states are destroyed [49,62], transforming into closed Fermi loops that disconnect from the bulk Dirac node projections [Fig. 1(c)]. This evolution of Fermi arc surface states, independent of the bulk Dirac nodes, challenges the conventional topological bulk-boundary correspondence and paves the way for exploring exotic topological phenomena. However, its impact on transport properties remains unclear, limiting further electronic applications.

Here we unveil the transition of surface Fermi arcs to closed Fermi loop in surface-modified $Cd_3As_2$ nanoplates through spin potentiometric and quantum oscillation experiments. The nanoplates are modified through the deposition of surface atoms or water molecules, intentionally introducing an interfacial electrical potential perturbation to the sample surface. It's found that the surface charge-spin conversion efficiency is enhanced as tuning the Fermi level into the conduction band. Additionally, magneto-transport measurements reveal dual-frequency surface Shubnikov-de Haas (SdH) oscillations with frequencies that increase linearly with gate voltage. These observations suggest the presence of closed surface Fermi loops, rather than conventional Fermi arcs.

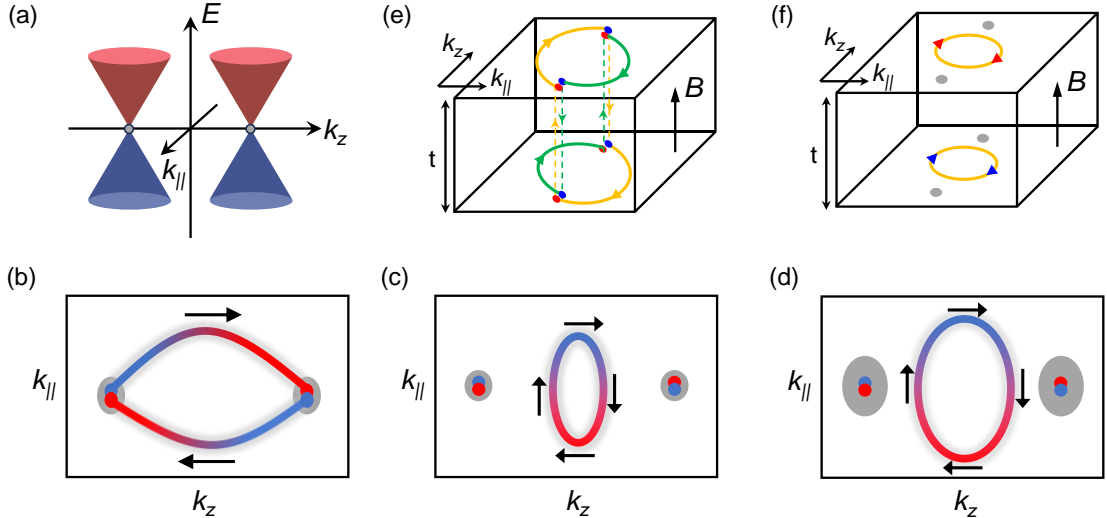

**FIG. 1. Fragile topology of surface states in the Dirac semimetal.**

(a) Schematic energy dispersion of Dirac semimetal $Cd_3As_2$ showing a pair of Dirac nodes along the $k_z$ axis.

(b) The expected double Fermi arcs on the surface Brillouin zones (BZs) of Dirac

semimetal Cd$_3$As$_2$. Red and blue denote the Weyl nodes with opposite chirality, and the black arrows represent the spins locked to the momentum. The $E_F = 0$ corresponds to the Fermi level situated at bulk Dirac nodes.

(c),(d) Deformation of double Fermi arcs into closed Fermi loop under a symmetry-breaking surface potential perturbation. As increasing the Fermi level from (c) $E_F = 0$ to (d) $E_F > 0$, the area of closed Fermi pocket is enlarged. The brown shaded regions denote the projections of bulk states onto the surface BZ. The black arrows represent the spins locked to the momentum.

(e) Schematic of Weyl orbits in Cd$_3$As$_2$ with a magnetic field $B$ perpendicular to the sample surface. Each set of Weyl orbit is composed of Fermi arcs in the momentum space on the two opposite surfaces and a real-space propagation through the bulk.

(f) Schematic of two Fermi loops on the sample surface, which individually constitutes a closed magnetic orbit on each surface.

We elaborate here on the experimental methods used to characterize surface state properties: spin potentiometric and quantum oscillation measurements (see Supplemental Material, Note 1 [63]). Spin potentiometric measurements [64-67] detect current-induced spin polarization in surface states, which is expected in both Fermi arc and Fermi loop surface states due to spin-momentum locking [Figs. 1(b)-1(d)]; however, they should exhibit distinct gate dependencies (see Supplemental Material, Note 2 [63]). Quantum oscillation measurements, in turn, provide information on the Fermi pocket area of surface states via oscillation frequency. In the case of a Weyl orbit formed by double Fermi arcs [Fig. 1(e)], the Fermi arcs on opposite surfaces are connected through bulk Weyl nodes, and the oscillation frequency scales linearly with the bulk Fermi wave vector $k_F$ [49,50]. For Fermi loop surface states [Fig. 1(f)], a closed magnetic orbit forms on a single surface, and its oscillation frequency is proportional to the square of the surface Fermi wave vector $k_F$. Experimentally, we can use gate voltage to modulate the bulk/surface Fermi wave vector and examine the gate dependence to distinguish between these two surface state scenarios. Additionally, Fermi loop surface states yield a π Berry phase shift in SdH oscillations, whereas for Weyl orbits, the bulk tunneling process introduces a sample-thickness-dependent phase term [49].

The $Cd_3As_2$ nanoplates were grown via chemical vapor deposition method [45,68], and individual nanoplate with the thickness 80~100 nm was transferred to the Si substrate with oxide layer on top. The Si substrate works as the back gate to modulate the Fermi level of nanoplate. For the devices used in the spin potentiometric study, Ti/Au and Co/Au electrodes were fabricated using two rounds of standard e-beam lithography and e-beam evaporation. As shown in Fig. 2(a), the outer Ti/Au electrodes serve as source and drain contacts to inject charge currents into the nanoplate, while the ferromagnetic Co electrode detects the current-induced spin polarization. After electrode fabrication, heavy metal atoms were deposited onto the nanoplate surface via e-beam evaporation. The quantity of heavy metal atoms was carefully controlled to avoid forming a continuous conductive layer, ensuring current predominantly flows through the nanoplate. In addition to surface atomic modifications, devices with adsorbed water molecules were also studied. By introducing a surface potential perturbation through atomic deposition or water molecule adsorption, deformation of surface Fermi arcs can be induced. This study focuses on devices A-F: device A is decorated with Cu atoms, devices B-D with Au atoms, and devices E and F with water molecule adsorption (see Supplemental Material, Table S1 [63]).

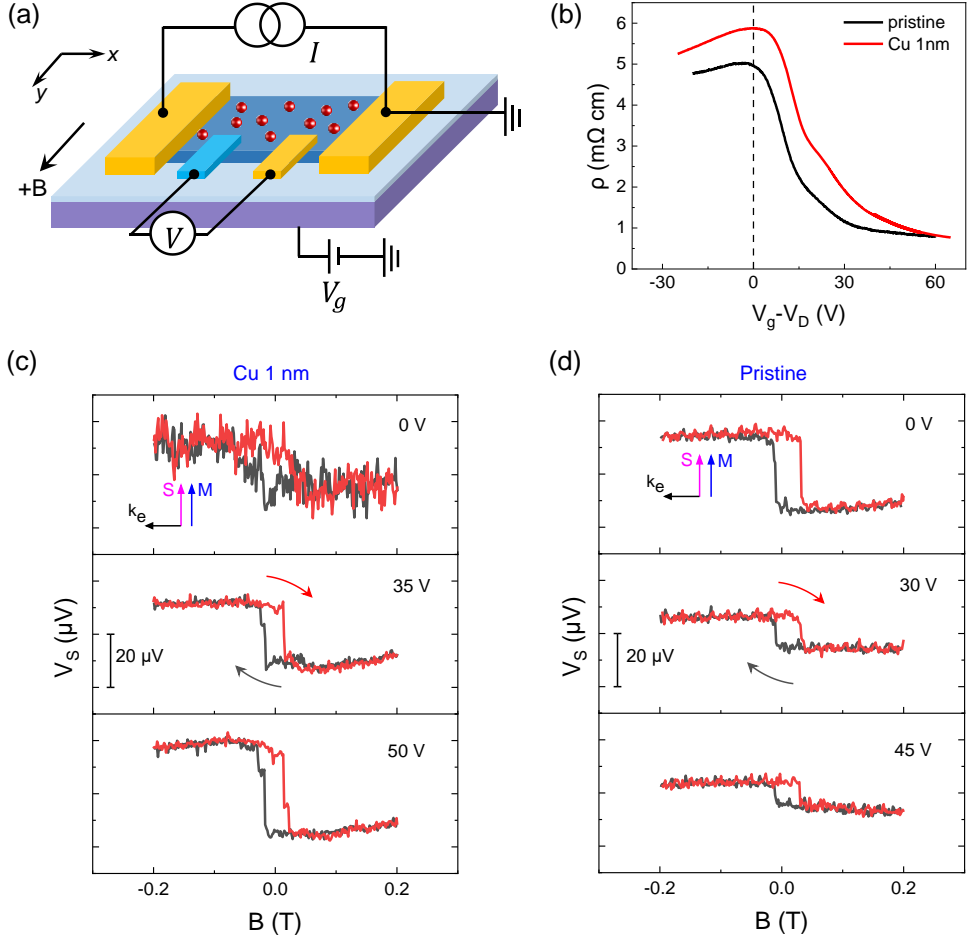

**FIG. 2. Enhanced current-induced spin polarization in the conduction band of Cd$_3$As$_2$ nanoplate decorated with Cu atoms.**

(a) Schematic of the device and measurement configuration. The Cd$_3$As$_2$ nanoplate is shown in blue, with red spheres representing metal atoms on its surface. Ti/Au and Co electrodes are shown in orange and light blue, respectively.

(b) Transfer curves of the surface-modified nanoplate (device A, decorated with 1 nm-thick Cu) and the pristine nanoplate (Ref. [68]), measured using standard four-probe geometry. Gate voltages are given relative to the Dirac point $V_D$, where $V_D = -20$ and -30 V for the surface-modified and pristine nanoplate, respectively.

(c) Spin-dependent voltage ($V_S$) hysteresis loops under $V_g - V_D = 0$, 35 and 50 V in device A, with an applied DC current of 30 μA.

(d) Magnetic hysteresis loops of the $V_S$ under $V_g - V_D = 0$, 30 and 45 V in the pristine nanoplate, with an applied DC current of 20 μA. The insets in (c) and (d) illustrate the electron spin locked to the momentum.

As illustrated in Fig. 2(a), we apply a direct current $I$ to the nanoplate device through the outer two Ti/Au electrodes and simultaneously measure the voltage $V$ between the inner Co and Ti/Au electrodes. The magnetic field is swept along the $y$-axis to modulate the magnetization $M$ of Co. The Cu-decorated nanoplate device A exhibits typical bipolar transport behavior, comparable to that of the pristine nanoplate [Fig. 2(b)]. The transfer curves reveal that the electron and hole mobilities of bulk states in device A are comparable to those in the pristine nanoplate, indicating similar bulk state characteristics. Figure 2(c) shows the measured spin-dependent voltage ($V_S$) versus magnetic field $B$ at different $V_g$ in device A. A magnetic hysteresis loop is observed near the Dirac point ($V_g - V_D = 0$), indicating the presence of net spin polarization ($S$) induced by the charge current. The $V_S$ is obtained by subtracting the spin-independent bulk signals from the raw data (see Supplemental Material, Fig. S2 [63]). According to the rule that $V_S > 0$ when $S \parallel M$, and $V_S < 0$ when $S \parallel -M$, the induced $S$ is along $-y$ direction [66,69]. Considering the net electron momentum ($k_e$) under current, $S$ is found locked to $k_e$ at right-handed angle [inset in Fig. 2(c)]. As the gate voltage increases, raising the Fermi level to 35 and 50 V above the Dirac point, the loop polarity remains unchanged but its amplitude shows an anomalous increase [Fig. 2(c)].

The window height of the magnetic loop ($\Delta V_S$), defined as the voltage difference at zero magnetic field for different sweeping directions, is widely used to quantify the amplitude of current-induced spin polarization and access the charge-spin conversion efficiency [44,45,64,65,68,70]. At $V_g = V_D + 50$ V, $\Delta V_S$ reaches 30 μV, nearly three times its value at $V_g = V_D$. In contrast, increasing the gate voltage in a pristine nanoplate significantly reduces $\Delta V_S$ [Fig. 2(d)], which has been widely observed in previous works [45,68]. The observed enhancement in $\Delta V_S$ for surface-modified device A cannot be explained by a Fermi arc surface state model, which would predict a decrease in surface current and spin polarization due to greater bulk involvement as the Fermi level shifts from the Dirac point (see Supplemental Material, Fig. S1 [63]).

Instead, the gate-enhanced $\Delta V_S$ aligns with a closed Fermi loop scenario [Figs. 1(c) and 1(d)], which has a higher surface density of states (DOS) at elevated Fermi levels, leading to anticipated increases in spin polarization (see Supplemental Material, Note 2 [63]).

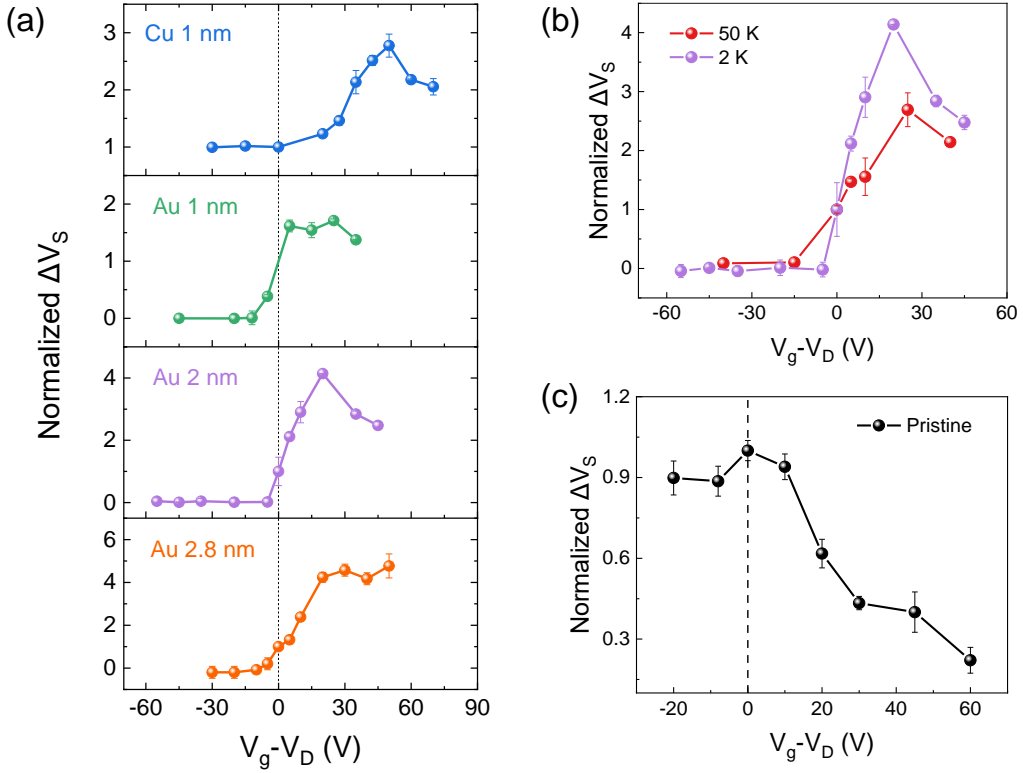

**FIG. 3. Gate voltage modulation of charge-to-spin conversion in surface-modified nanoplates.**

(a) Normalized $V_S$ loop height, *i.e.*, $\Delta V_S(V_g - V_D) / \Delta V_S(V_g - V_D = 0)$, measured in device A (decorated with Cu atoms) and devices B-D (decorated with Au atoms).

(b) Gate-tuned spin polarization on/off effect in device C (decorated with 2 nm-thick Au) under 2 and 50 K.

(c) Gate voltage dependence of normalized $\Delta V_S$ in the pristine nanoplate. Error bars represent the standard deviations over multiple measurements in panels (a)-(c).

We further tune the Fermi level deeply into both the conduction and valence bands to modulate the spin polarization signal in various devices, as shown in Fig. 3. To capture the variation of $\Delta V_S$ with gate voltage, the $\Delta V_S$ is normalized by $\Delta V_S(V_g - $ $V_D = 0$). As shown in the uppermost panel of Fig. 3(a) for device A, the spin polarization amplitude increases steadily up to $V_g - V_D = 50$ V, after which it slightly decreases with further gate voltage increases. A similar behavior is observed in nanoplates decorated with Au atoms (devices B–D, lower panels in Fig. 3(a)) and in a nanoplate adsorbed with water molecules (see device E in Supplemental Material, Fig. S4 [63]).

As lifting the Fermi level above the Dirac point, the amplitude of spin polarization signals shows an overall increase. In contrast, within the valence band, the spin polarization signals drop sharply, even quenching to zero. The gate-tuned on/off ratio, observed in device C, reaches up to 47000%, which is two orders of magnitude larger than previously reported results [71,72]. Notably, gate-switchable spin polarization signals are observed at 50 K [Fig. 3(b)]. The response of spin polarization signals to the gate voltage in surface-modified nanoplates differs significantly from that in pristine nanoplates [Fig. 3(c)], further indicating a disruption of Fermi arc surface states. This gate-enhanced $\Delta V_S$ likely originates from the enlarged Fermi pocket and increased DOS associated with Fermi loop surface states. Below the Dirac point, the reduction of $\Delta V_S$ may result from a decline in surface spin polarization due to intense hole scattering in the valence band [73,74].

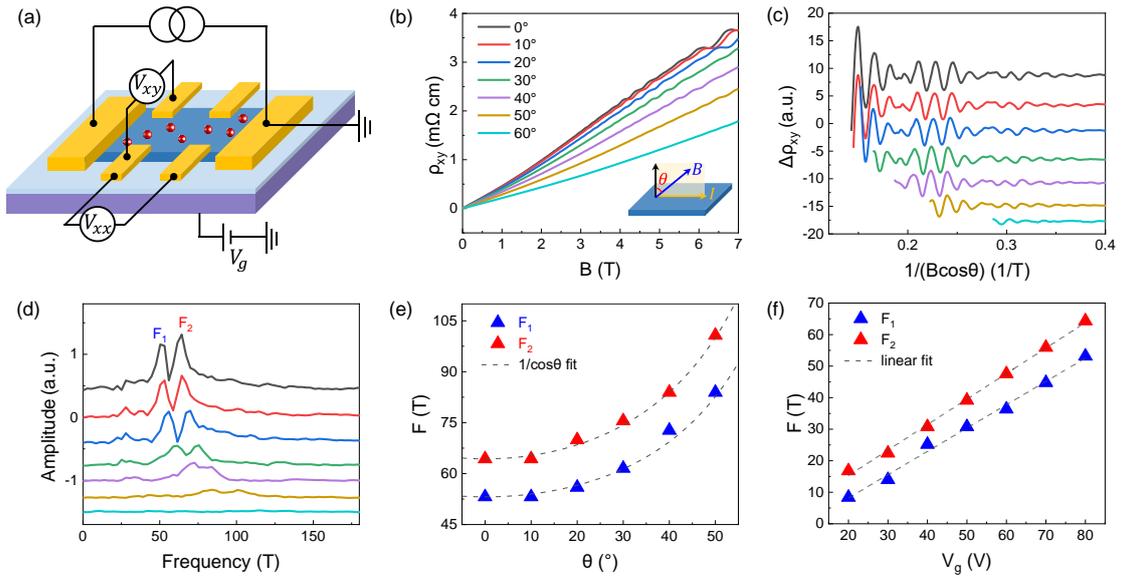

**FIG. 4. Dual-frequency SdH oscillations from surface states in a Cd$_3$As$_2$ nanoplate**

**with adsorbed water molecules.**

(a) Schematic of the device measurement configuration.

(b) The Hall resistivity ($\rho_{xy}$) under magnetic fields tilting at different angles ($\theta$) in device F. $\theta$ is the angle between the magnetic field and the surface normal of the nanoplate, as indicated in the inset. The gate voltage is set to 85 V above the Dirac point, $V_D$.

(c) SdH oscillations versus $1/(B\cos\theta)$ at different $\theta$ values, extracted from the data in (b).

(d) FFT spectra of SdH oscillations in (c), revealing two primary frequencies, $F_1$ and $F_2$.

(e) Angular dependence of $F_1$ and $F_2$ obtained from FFT analysis. Dashed lines represent the $1/\cos\theta$ fits.

(f) Gate voltage dependence of $F_1$ and $F_2$. Dashed lines denote the linear fits.

The deformation of surface Fermi arcs has also been unveiled in quantum oscillations experiments on device F, where intentional adsorption of gas molecules (*e.g.*, H₂O) was achieved by exposing the cooled device to ambient air. (see Supplemental Material, Note 1 [63]). Figure 4 presents the angle- and gate voltage-dependent SdH oscillations in device F. The longitudinal resistivity $\rho_{xx}$ and the Hall resistivity $\rho_{xy}$ were measured simultaneously by a standard six-probe method [Fig. 4(a)] by sweeping the magnetic field. SdH oscillations appear in both $\rho_{xx}$ and $\rho_{xy}$, with the latter showing more pronounced oscillations (see Supplemental Material, Fig. S5 [63]). Besides, the maxima in $\rho_{xy}$ are well consistent with the minima in conductivity $\sigma_{xx}$ (see Supplemental Material, Fig. S6 [63]).

Figure 4(b) plots $\rho_{xy}$ versus $B$ for different angles $\theta$ between $B$ and the nanoplate's normal direction [inset of Fig. 4(b)]. Oscillations rapidly attenuate with increasing $\theta$ and become indistinct above 50°. After background subtraction, $\Delta\rho_{xy}$ is plotted against $1/(B\cos\theta)$ to capture the effect of the out-of-plane component of the magnetic field [Fig. 4(c)]. Two key features emerge: first, the oscillations depend solely on $B\cos\theta$ across angles; second, a clear beating pattern suggests two oscillatory components. Fast Fourier-transform (FFT) analysis identifies two main frequencies, $F_1$ and $F_2$ [Fig. 4(d)], whose angular dependence follows a $1/\cos\theta$ relation [dashed lines in Fig. 4(e)], indicating two-dimensional SdH oscillations from surface states of the nanoplate.

To further explore the surface states, we analyzed the gate voltage dependence of SdH oscillations (see Supplemental Material, Figs. S7 and S8 [63]). Figure 4(f) shows the linear dependence of $F_1$ and $F_2$ on gate voltage, contradicting the expected $V_g^{1/3}$ scaling for a Dirac semimetal's Fermi arc state [49,50,53,75] and consistent with Fermi loop surface states [76,77], where frequency scales with $k_F^2$, carrier density n, or $V_g$ (see Supplemental Material, Note 7 [63]). Landau fan analysis indicates intercepts of -0.41 and -0.55 for the two oscillations (see Supplemental Material, Fig. S9 [63]), aligning with a Berry phase of π, as expected for spin-helical Fermi loop surface states [78-80].

In summary, we have demonstrated the evolution from Fermi arc to Fermi loop surface states in modified Dirac semimetal nanoplates through current-induced spin polarization and SdH oscillation experiments. Our findings show that depositing heavy metal atoms or adsorbing water molecules on the nanoplate surface enhances the charge-to-spin conversion efficiency and enables gate-switching control. The SdH oscillations reveal the two-dimensional transport properties and topological nature of Fermi loop states, with a Berry phase of π. These results highlight the sensitivity of surface states to modifications in Dirac semimetals, uncovering the fragile topology of Fermi arc surface states. The Fermi loop topological surface states in modified Dirac semimetals hold promise for future spintronic and electronic devices.

**Acknowledgements**
This work was supported by the National Natural Science Foundation of China (Grant Nos. 62425401, 62321004 and 12204016), and Innovation Program for Quantum Science and Technology (Grant No. 2021ZD0302403). C.Li acknowledges Dutch Research Council (NWO) for the financial support of the project SuperHOTS with file

number VI.Vidi.203.047.

# Reference

[1] A. A. Burkov, Topological semimetals, *Nat. Mater.* **15**, 1145 (2016).

[2] S. Wang, B.-C. Lin, A.-Q. Wang, D.-P. Yu, and Z.-M. Liao, Quantum transport in Dirac and Weyl semimetals: a review, *Adv. Phys.: X* **2**, 518 (2017).

[3] N. P. Armitage, E. J. Mele, and A. Vishwanath, Weyl and Dirac semimetals in three-dimensional solids, *Rev. Mod. Phys.* **90**, 015001 (2018).

[4] A.-Q. Wang, X.-G. Ye, D.-P. Yu, and Z.-M. Liao, Topological Semimetal Nanostructures: From Properties to Topotronics, *ACS Nano* **14**, 3755 (2020).

[5] P. Z. Liu, J. R. Wilhams, and J. J. Cha, Topological nanomaterials, *Nat. Rev. Mater.* **4**, 479 (2019).

[6] Z. Wang, Y. Sun, X.-Q. Chen, C. Franchini, G. Xu, H. Weng, X. Dai, and Z. Fang, Dirac semimetal and topological phase transitions in $A_3$Bi (A=Na, K, Rb), *Phys. Rev. B* **85**, 195320 (2012).

[7] Z. Wang, H. Weng, Q. Wu, X. Dai, and Z. Fang, Three-dimensional Dirac semimetal and quantum transport in $Cd_3As_2$, *Phys. Rev. B* **88**, 125427 (2013).

[8] Z. K. Liu, B. Zhou, Y. Zhang, Z. J. Wang, H. M. Weng, D. Prabhakaran, S.-K. Mo, Z. X. Shen, Z. Fang, X. Dai *et al.*, Discovery of a Three-Dimensional Topological Dirac Semimetal $Na_3Bi$, *Science* **343**, 864 (2014).

[9] H. Yi, Z. Wang, C. Chen, Y. Shi, Y. Feng, A. Liang, Z. Xie, S. He, J. He, Y. Peng *et al.*, Evidence of Topological Surface State in Three-Dimensional Dirac Semimetal $Cd_3As_2$, *Sci. Rep.* **4**, 6106 (2014).

[10] M. Neupane, S. Y. Xu, R. Sankar, N. Alidoust, G. Bian, C. Liu, I. Belopolski, T. R. Chang, H. T. Jeng, H. Lin *et al.*, Observation of a three-dimensional topological Dirac semimetal phase in high-mobility $Cd_3As_2$, *Nat. Commun.* **5**, 3786 (2014).

[11] S.-Y. Xu, C. Liu, S. K. Kushwaha, R. Sankar, J. W. Krizan, I. Belopolski, M. Neupane, G. Bian, N. Alidoust, T.-R. Chang *et al.*, Observation of Fermi arc surface states in a topological metal, *Science* **347**, 294 (2015).

[12] S. Jeon, B. B. Zhou, A. Gyenis, B. E. Feldman, I. Kimchi, A. C. Potter, Q. D. Gibson, R. J. Cava, A. Vishwanath, and A. Yazdani, Landau quantization and quasiparticle interference in the three-dimensional Dirac semimetal $Cd_3As_2$, *Nat. Mater.* **13**, 851 (2014).

[13] C. Z. Li, L. X. Wang, H. Liu, J. Wang, Z. M. Liao, and D. P. Yu, Giant negative magnetoresistance induced by the chiral anomaly in individual $Cd_3As_2$ nanowires, *Nat. Commun.* **6**, 10137 (2015).

[14] J. Xiong, S. K. Kushwaha, T. Liang, J. W. Krizan, M. Hirschberger, W. Wang, R. J. Cava, and N. P. Ong, Evidence for the chiral anomaly in the Dirac semimetal $Na_3Bi$, *Science* **350**, 413 (2015).

[15] L. P. He, X. C. Hong, J. K. Dong, J. Pan, Z. Zhang, J. Zhang, and S. Y. Li, Quantum Transport Evidence for the Three-Dimensional Dirac Semimetal Phase in $Cd_3As_2$, *Phys. Rev. Lett.* **113**, 246402 (2014).

[16] Y. Zhao, H. Liu, C. Zhang, H. Wang, J. Wang, Z. Lin, Y. Xing, H. Lu, J. Liu, Y. Wang *et al.*, Anisotropic Fermi Surface and Quantum Limit Transport in High Mobility Three-Dimensional Dirac Semimetal $Cd_3As_2$, *Phys. Rev. X* **5**, 031037 (2015).

[17] S. M. Young, S. Zaheer, J. C. Y. Teo, C. L. Kane, E. J. Mele, and A. M. Rappe, Dirac Semimetal in Three Dimensions, *Phys. Rev. Lett.* **108**, 140405 (2012).

[18] B.-J. Yang and N. Nagaosa, Classification of stable three-dimensional Dirac semimetals with nontrivial topology, *Nat. Commun.* **5**, 4898 (2014).

[19] E. V. Gorbar, V. A. Miransky, and I. A. Shovkovy, Engineering Weyl nodes in Dirac semimetals by a magnetic field, *Phys. Rev. B* **88**, 165105 (2013).

[20] B. C. Lin, S. Wang, S. Wiedmann, J. M. Lu, W. Z. Zheng, D. Yu, and Z. M. Liao, Observation of an Odd-Integer Quantum Hall Effect from Topological Surface States in $Cd_3As_2$, *Phys. Rev. Lett.* **122**, 036602 (2019).

[21] W.-Z. Zheng, X.-G. Ye, B.-C. Lin, R.-R. Li, D.-P. Yu, and Z.-M. Liao, Magnetotransport evidence for topological phase transition in a Dirac semimetal, *Appl. Phys. Lett.* **115** (2019).

[22] J. L. Collins, A. Tadich, W. Wu, L. C. Gomes, J. N. B. Rodrigues, C. Liu, J. Hellerstedt, H. Ryu, S. Tang, S.-K. Mo *et al.*, Electric-field-tuned topological phase transition in ultrathin $Na_3Bi$, *Nature* **564**, 390 (2018).

[23] H. Xia, Y. Li, M. Cai, L. Qin, N. Zou, L. Peng, W. Duan, Y. Xu, W. Zhang, and Y.-S. Fu, Dimensional Crossover and Topological Phase Transition in Dirac Semimetal $Na_3Bi$ Films, *ACS Nano* **13**, 9647 (2019).

[24] L. Aggarwal, A. Gaurav, G. S. Thakur, Z. Haque, A. K. Ganguli, and G. Sheet, Unconventional superconductivity at mesoscopic point contacts on the 3D Dirac semimetal $Cd_3As_2$, *Nat. Mater.* **15**, 32 (2016).

[25] H. Wang, H. Wang, H. Liu, H. Lu, W. Yang, S. Jia, X.-J. Liu, X. C. Xie, J. Wei, and J. Wang, Observation of superconductivity induced by a point contact on 3D Dirac semimetal $Cd_3As_2$ crystals, *Nat. Mater.* **15**, 38 (2016).

[26] L. He, Y. Jia, S. Zhang, X. Hong, C. Jin, and S. Li, Pressure-induced superconductivity in the three-dimensional topological Dirac semimetal $Cd_3As_2$, *npj Quantum Mater.* **1**, 16014 (2016).

[27] X. Cheng, R. Li, D. Li, Y. Li, and X.-Q. Chen, Combined fast reversible liquidlike elastic deformation with topological phase transition in $Na_3Bi$, *Phys. Rev. B* **92**, 155109 (2015).

[28] W.-Z. Zheng, T.-Y. Zhao, A.-Q. Wang, D.-Y. Xu, P.-Z. Xiang, X.-G. Ye, and Z.-M. Liao, Strain-gradient induced topological transition in bent nanoribbons of the Dirac semimetal $Cd_3As_2$, *Phys. Rev. B* **104**, 155140 (2021).

[29] D. Shao, J. Ruan, J. Wu, T. Chen, Z. Guo, H. Zhang, J. Sun, L. Sheng, and D. Xing, Strain-induced quantum topological phase transitions in $Na_3Bi$, *Phys. Rev. B* **96**, 075112 (2017).

[30] N. P. Armitage, E. J. Mele, and A. Vishwanath, Weyl and Dirac semimetals in three-dimensional solids, *Rev. Mod. Phys.* **90**, 015001 (2018).

[31] B. Q. Lv, T. Qian, and H. Ding, Experimental perspective on three-dimensional topological semimetals, *Rev. Mod. Phys.* **93**, 025002 (2021).

[32] C. Zhang, Y. Zhang, H. Z. Lu, X. C. Xie, and F. X. Xiu, Cycling Fermi arc electrons with Weyl orbits, *Nat. Rev. Phys.* **3**, 660 (2021).

[33] S. Wang, B. C. Lin, A. Q. Wang, D. P. Yu, and Z. M. Liao, Quantum transport in Dirac and Weyl semimetals: a review, *Adv. Phys.: X* **2**, 518 (2017).

[34] X. Wan, A. M. Turner, A. Vishwanath, and S. Y. Savrasov, Topological semimetal and Fermi-arc surface states in the electronic structure of pyrochlore iridates, *Phys. Rev. B* **83**, 205101 (2011).

[35] S.-Y. Xu, I. Belopolski, N. Alidoust, M. Neupane, G. Bian, C. Zhang, R. Sankar, G. Chang, Z. Yuan, C.-C. Lee *et al.*, Discovery of a Weyl fermion semimetal and topological Fermi arcs, *Science* **349**, 613 (2015).

[36] H. Weng, C. Fang, Z. Fang, B. A. Bernevig, and X. Dai, Weyl Semimetal Phase in Noncentrosymmetric Transition-Metal Monophosphides, *Phys. Rev. X* **5**, 011029 (2015).

[37] S.-M. Huang, S.-Y. Xu, I. Belopolski, C.-C. Lee, G. Chang, B. Wang, N. Alidoust, G. Bian, M. Neupane, C. Zhang *et al.*, A Weyl Fermion semimetal with surface Fermi arcs in the transition metal monopnictide TaAs class, *Nat. Commun.* **6**, 7373 (2015).

[38] B. Q. Lv, H. M. Weng, B. B. Fu, X. P. Wang, H. Miao, J. Ma, P. Richard, X. C. Huang, L. X. Zhao, G. F. Chen *et al.*, Experimental Discovery of Weyl Semimetal TaAs, *Phys. Rev. X* **5**, 031013 (2015).

[39] S.-Y. Xu, N. Alidoust, I. Belopolski, Z. Yuan, G. Bian, T.-R. Chang, H. Zheng, V. N. Strocov, D. S. Sanchez, G. Chang *et al.*, Discovery of a Weyl fermion state with Fermi arcs in niobium arsenide, *Nat. Phys.* **11**, 748 (2015).

[40] B. Q. Lv, N. Xu, H. M. Weng, J. Z. Ma, P. Richard, X. C. Huang, L. X. Zhao, G. F. Chen, C. E. Matt, F. Bisti *et al.*, Observation of Weyl nodes in TaAs, *Nat. Phys.* **11**, 724 (2015).

[41] E. V. Gorbar, V. A. Miransky, I. A. Shovkovy, and P. O. Sukhachov, Surface Fermi arcs in $Z_2$ Weyl semimetals $A_3$Bi (A=Na, K, Rb), *Phys. Rev. B* **91**, 235138 (2015).

[42] L.-X. Wang, C.-Z. Li, D.-P. Yu, and Z.-M. Liao, Aharonov-Bohm oscillations in Dirac semimetal $Cd_3As_2$ nanowires, *Nat. Commun.* **7**, 10769 (2016).

[43] B.-C. Lin, S. Wang, L.-X. Wang, C.-Z. Li, J.-G. Li, D. Yu, and Z.-M. Liao, Gate-tuned Aharonov-Bohm interference of surface states in a quasiballistic Dirac semimetal nanowire, *Phys. Rev. B* **95**, 235436 (2017).

[44] B.-C. Lin, S. Wang, A.-Q. Wang, Y. Li, R.-R. Li, K. Xia, D. Yu, and Z.-M. Liao, Electric Control of Fermi Arc Spin Transport in Individual Topological Semimetal Nanowires, *Phys. Rev. Lett.* **124**, 116802 (2020).

[45] A.-Q. Wang, P.-Z. Xiang, X.-G. Ye, W.-Z. Zheng, D. Yu, and Z.-M. Liao, Surface Engineering of Antisymmetric Linear Magnetoresistance and Spin-Polarized Surface State Transport in Dirac Semimetals, *Nano Lett.* **21**, 2026 (2021).

[46] A.-Q. Wang, P.-Z. Xiang, X.-G. Ye, W.-Z. Zheng, D. Yu, and Z.-M. Liao, Room-Temperature Manipulation of Spin Texture in a Dirac Semimetal, *Phys. Rev. Appl.* **14**, 054044 (2020).

[47] C.-Z. Li, A.-Q. Wang, C. Li, W.-Z. Zheng, A. Brinkman, D.-P. Yu, and Z.-M. Liao, Fermi-arc supercurrent oscillations in Dirac semimetal Josephson junctions, *Nat. Commun.* **11**, 1150 (2020).

[48] C.-Z. Li, A.-Q. Wang, C. Li, W.-Z. Zheng, A. Brinkman, D.-P. Yu, and Z.-M. Liao, Topological Transition of Superconductivity in Dirac Semimetal Nanowire Josephson Junctions, *Phys. Rev. Lett.* **126**, 027001 (2021).

[49] A. C. Potter, I. Kimchi, and A. Vishwanath, Quantum oscillations from surface Fermi arcs in Weyl and Dirac semimetals, *Nat. Commun.* **5**, 5161 (2014).

[50] P. J. W. Moll, N. L. Nair, T. Helm, A. C. Potter, I. Kimchi, A. Vishwanath, and J. G. Analytis, Transport evidence for Fermi-arc-mediated chirality transfer in the Dirac semimetal $Cd_3As_2$, *Nature* **535**, 266 (2016).

[51] Y. Zhang, D. Bulmash, P. Hosur, A. C. Potter, and A. Vishwanath, Quantum oscillations from generic surface Fermi arcs and bulk chiral modes in Weyl semimetals, *Sci. Rep.* **6**, 23741 (2016).

[52] C. Zhang, A. Narayan, S. H. Lu, J. L. Zhang, H. Q. Zhang, Z. L. Ni, X. Yuan, Y. W. Liu, J. H. Park, E. Z. Zhang *et al.*, Evolution of Weyl orbit and quantum Hall effect in Dirac semimetal $Cd_3As_2$, *Nat. Commun.* **8**, 1272 (2017).

[53] Y. Miyazaki, T. Yokouchi, K. Shibata, Y. Chen, H. Arisawa, T. Mizoguchi, E. Saitoh, and Y. Shiomi, Quantum oscillations from Fermi arc surface states in $Cd_3As_2$ submicron wires, *Phys. Rev. Res.* **4**, L022002 (2022).

[54] M. Uchida, Y. Nakazawa, S. Nishihaya, K. Akiba, M. Kriener, Y. Kozuka, A. Miyake, Y. Taguchi, M. Tokunaga, N. Nagaosa *et al.*, Quantum Hall states observed in thin films of Dirac semimetal $Cd_3As_2$, *Nat. Commun.* **8**, 2274 (2017).

[55] T. Schumann, L. Galletti, D. A. Kealhofer, H. Kim, M. Goyal, and S. Stemmer, Observation of the Quantum Hall Effect in Confined Films of the Three-Dimensional Dirac Semimetal $Cd_3As_2$, *Phys. Rev. Lett.* **120**, 016801 (2018).

[56] S. Nishihaya, M. Uchida, Y. Nakazawa, M. Kriener, Y. Kozuka, Y. Taguchi, and M. Kawasaki, Gate-tuned quantum Hall states in Dirac semimetal $(Cd_{1-x}Zn_x)_3As_2$, *Sci. Adv.* **4**, eaar5668 (2018).

[57] C. Zhang, Y. Zhang, X. Yuan, S. Lu, J. Zhang, A. Narayan, Y. Liu, H. Zhang, Z. Ni, R. Liu *et al.*, Quantum Hall effect based on Weyl orbits in $Cd_3As_2$, *Nature* **565**, 331 (2019).

[58] C. M. Wang, H. P. Sun, H. Z. Lu, and X. C. Xie, 3D Quantum Hall Effect of Fermi Arc in Topological Semimetals, *Phys. Rev. Lett.* **119**, 136806 (2017).

[59] M. Kargarian, M. Randeria, and Y.-M. Lu, Are the surface Fermi arcs in Dirac semimetals topologically protected?, *Proc. Natl. Acad. Sci. U. S. A.* **113**, 8648 (2016).

[60] M. Kargarian, Y.-M. Lu, and M. Randeria, Deformation and stability of surface states in Dirac semimetals, *Phys. Rev. B* **97**, 165129 (2018).

[61] Y. Wu, N. H. Jo, L.-L. Wang, C. A. Schmidt, K. M. Neilson, B. Schrunk, P. Swatek, A. Eaton, S. L. Bud'ko, P. C. Canfield *et al.*, Fragility of Fermi arcs in Dirac semimetals, *Phys. Rev. B* **99**, 161113 (2019).

[62] C. Le, X. Wu, S. Qin, Y. Li, R. Thomale, F.-C. Zhang, and J. Hu, Dirac semimetal in *β*-CuI without surface Fermi arcs, *Proc. Natl. Acad. Sci. U. S. A.*

**115**, 8311 (2018).

[63] See Supplemental Material for surface modification and transport measurement, discussion on the gate-tunable surface spin polarization signals, spin potentiometric measurement results in different devices, gate voltage dependence of SdH oscillations, analysis on the Landau fan diaagram, and comparison of three types of magentic orbits.

[64] C. H. Li, O. M. J. van 't Erve, J. T. Robinson, Y. Liu, L. Li, and B. T. Jonker, Electrical detection of charge-current-induced spin polarization due to spin-momentum locking in $Bi_2Se_3$, *Nat. Nanotechnol.* **9**, 218 (2014).

[65] A. Dankert, J. Geurs, M. V. Kamalakar, S. Charpentier, and S. P. Dash, Room Temperature Electrical Detection of Spin Polarized Currents in Topological Insulators, *Nano Lett.* **15**, 7976 (2015).

[66] S. Hong, V. Diep, S. Datta, and Y. P. Chen, Modeling potentiometric measurements in topological insulators including parallel channels, *Phys. Rev. B* **86**, 085131 (2012).

[67] J. A. Voerman, C. Li, Y. Huang, and A. Brinkman, Spin-Momentum Locking in the Gate Tunable Topological Insulator $BiSbTeSe_2$ in Non-Local Transport Measurements, *Adv. Electron. Mater.* **5**, 1900334 (2019).

[68] A.-Q. Wang, P.-Z. Xiang, X.-G. Ye, W.-Z. Zheng, D. Yu, and Z.-M. Liao, Room-Temperature Manipulation of Spin Texture in a Dirac Semimetal, *Phys. Rev. Appl.* **14**, 054044 (2020).

[69] F. Yang, S. Ghatak, A. A. Taskin, K. Segawa, Y. Ando, M. Shiraishi, Y. Kanai, K. Matsumoto, A. Rosch, and Y. Ando, Switching of charge-current-induced spin polarization in the topological insulator $BiSbTeSe_2$, *Phys. Rev. B* **94**, 075304 (2016).

[70] A.-Q. Wang, P.-Z. Xiang, T.-Y. Zhao, and Z.-M. Liao, Topological nature of higher-order hinge states revealed by spin transport, *Sci. Bull.* **67**, 788 (2022).

[71] P. Chuang, S.-C. Ho, L. W. Smith, F. Sfigakis, M. Pepper, C.-H. Chen, J.-C. Fan, J. P. Griffiths, I. Farrer, H. E. Beere *et al.*, All-electric all-semiconductor spin field-effect transistors, *Nat. Nanotechnol.* **10**, 35 (2015).

[72] M. Cahay, Closer to an all-electric device, *Nat. Nanotechnol.* **10**, 21 (2015).

[73] Z. K. Liu, J. Jiang, B. Zhou, Z. J. Wang, Y. Zhang, H. M. Weng, D. Prabhakaran, S. K. Mo, H. Peng, P. Dudin *et al.*, A stable three-dimensional topological Dirac semimetal $Cd_3As_2$, *Nat. Mater.* **13**, 677 (2014).

[74] M. Neupane, S.-Y. Xu, R. Sankar, N. Alidoust, G. Bian, C. Liu, I. Belopolski, T.-R. Chang, H.-T. Jeng, H. Lin *et al.*, Observation of a three-dimensional topological Dirac semimetal phase in high-mobility $Cd_3As_2$, *Nat. Commun.* **5**, 3786 (2014).

[75] T. Liang, Q. Gibson, M. N. Ali, M. Liu, R. J. Cava, and N. P. Ong, Ultrahigh mobility and giant magnetoresistance in the Dirac semimetal $Cd_3As_2$, *Nat. Mater.* **14**, 280 (2015).

[76] L. A. Jauregui, M. T. Pettes, L. P. Rokhinson, L. Shi, and Y. P. Chen, Gate Tunable Relativistic Mass and Berry's phase in Topological Insulator Nanoribbon Field Effect Devices, *Sci. Rep.* **5**, 8452 (2015).

[77] P. Leng, F. Chen, X. Cao, Y. Wang, C. Huang, X. Sun, Y. Yang, J. Zhou, X. Xie, Z. Li *et al.*, Gate-Tunable Surface States in Topological Insulator β-$Ag_2$Te with High Mobility, *Nano Lett.* **20**, 7004 (2020).

[78] Z. Ren, A. A. Taskin, S. Sasaki, K. Segawa, and Y. Ando, Large bulk resistivity and surface quantum oscillations in the topological insulator $Bi_2Te_2Se$, *Phys. Rev. B* **82**, 241306 (2010).

[79] D.-X. Qu, Y. S. Hor, J. Xiong, R. J. Cava, and N. P. Ong, Quantum Oscillations and Hall Anomaly of Surface States in the Topological Insulator $Bi_2Te_3$, *Science* **329**, 821 (2010).

[80] J. G. Analytis, R. D. McDonald, S. C. Riggs, J.-H. Chu, G. S. Boebinger, and I. R. Fisher, Two-dimensional surface state in the quantum limit of a topological insulator, *Nat. Phys.* **6**, 960 (2010).